\documentclass[12pt,journal,final,onecolumn]{IEEEtran} 

\usepackage[pdftex]{graphicx}
\usepackage[cmex10]{amsmath}
\usepackage{amssymb}
\usepackage{amsthm}
\usepackage{amsfonts}
\usepackage{bm}
\usepackage{cite}
\usepackage[svgnames]{xcolor} 
\usepackage{pstricks,pst-node,pst-plot,pstricks-add}
\usepackage[tight,footnotesize]{subfigure}
\usepackage{algorithm}
\usepackage{algpseudocode}
\usepackage{hyperref} 

\graphicspath{{figs/}}

\input{niesen.def}

\newcommand{\Rs}{R_\msf{S}}
\newcommand{\RE}{R_\msf{E}}
\newcommand{\Rcac}{R_\msf{CAC}}
\newcommand{\Recc}{R_\msf{ECC}}
\newcommand{\Rldpc}{R_\msf{LDPC}}

\newcommand{\de}{d_\msf{E}}
\newcommand{\decc}{d_\msf{ECC}}

\newcommand{\Ccac}{\mc{C}_\msf{CAC}}
\newcommand{\Cecc}{\mc{C}_\msf{ECC}}

\newcommand{\xcac}{x_\msf{CAC}}
\newcommand{\ycac}{y_\msf{CAC}}
\newcommand{\xecc}{x_\msf{ECC}}
\newcommand{\yecc}{y_\msf{ECC}}
\newcommand{\xp}{x_\msf{P}}
\newcommand{\yp}{y_\msf{P}}

\begin{document}

\title{Joint Crosstalk-Avoidance and Error-Correction Coding for Parallel Data Buses} 

\author{Urs Niesen and Shrinivas Kudekar%
    \thanks{The authors are with Qualcomm's New Jersey Research Center,
        Bridgewater, NJ 08807. 
        Emails: \{uniesen,skudekar\}@qti.qualcomm.com}%
}


\maketitle

\begin{abstract}
Decreasing transistor sizes and lower voltage swings cause two distinct
problems for communication in integrated circuits.  First, decreasing inter-wire
spacing increases interline capacitive coupling, which adversely affects
transmission energy and delay. Second, lower voltage swings render the
transmission susceptible to various noise sources.  Coding can be used to
address both these problems. So-called crosstalk-avoidance codes mitigate
capacitive coupling, and traditional error-correction codes introduce
resilience against channel errors. 

Unfortunately, crosstalk-avoidance and error-correction codes cannot be combined
in a straightforward manner. On the one hand, crosstalk-avoidance encoding
followed by error-correction encoding destroys the crosstalk-avoidance property.
On the other hand, error-correction encoding followed by crosstalk-avoidance
encoding causes the crosstalk-avoidance decoder to fail in the presence of
errors. Existing approaches circumvent this difficulty by using additional bus
wires to protect the parities generated from the output of the error-correction
encoder, and are therefore inefficient. 

In this work we propose a novel joint crosstalk-avoidance and error-correction
coding and decoding scheme that provides higher bus transmission rates compared
to existing approaches. Our joint approach carefully embeds the parities such
that the crosstalk-avoidance property is preserved. We analyze the rate and
minimum distance of the proposed scheme. We also provide a density evolution
analysis and predict iterative decoding thresholds for reliable communication
under random bus erasures. This density evolution analysis is nonstandard, since
the crosstalk-avoidance constraints are inherently nonlinear. 
\end{abstract}

\section{Introduction} 
\label{sec:intro}

\subsection{Motivation and Related Work}
\label{sec:intro_motivation}

The various components in an integrated circuit communicate with each other via
parallel metal wires referred to as \emph{interconnects} or \emph{parallel buses}
\cite{Micheli02, Micheli10, Shanbag04, FutureofWires}.  As the inter-wire
spacing decreases with technology scaling, the coupling capacitance between
adjacent wires increases, leading to increased crosstalk on the bus
\cite{SotiriadisThesis}. Moreover, scaling of the supply voltage renders bus
transmissions susceptible to random errors. Together, these two effects increase
transmission energy/delay and degrade signal integrity. As a consequence, they
pose a major challenge to keep up with the demand for increasing data transfer
rates over interconnects.

Crosstalk-avoidance coding (CAC) encodes data for transmission over a
capacitively-coupled interconnect with reduced energy consumption and
propagation delay. This is achieved, for example, by avoiding opposing bit transitions on
adjacent bus wires \cite{SotiriadisThesis, Victor01, KonstantakopoulosThesis,
Chang10}.  

Error-correction coding (ECC) encodes data by adding redundancy in the form of
parity bits. This redundancy can then be used to detect and correct errors
occurring during the transmission. Such error correction is becoming
increasingly important for communication over interconnects. For example, the
DDR4 SDRAM standard uses an $8$-bit cyclic-redundancy code to provide error
correction for $72$ bits of data transmitted over the bus \cite{DDR4}. Thus, the
total number of bus wires used is $80$.

Since parallel buses benefit from the use of both CAC and ECC, it is beneficial
to try to combine them. However, this turns out to be nontrivial. The two
natural approaches, which use an ECC inner code followed by a CAC outer code, or
which use a CAC inner code followed by an ECC outer code both fail (the former
because the CAC outer code cannot be reliably decoded in the presence of errors,
the latter since the ECC outer code destroys the crosstalk-avoidance property).
Thus, crosstalk-avoidance and error-correction coding need to be performed
jointly.

The best known way to combine CAC and ECC is \cite{Sridhara05, Sridhara07}. They
propose to first encode the information using a CAC. Next an ECC is used to
generate parity bits for the output of the CAC. Each of these parity bits is
transmitted using two wires to ensure crosstalk avoidance. As a result of this
duplication, this approach is inefficient. For example, using the above DDR4
numbers, if the $72$ bits are already CAC encoded, then the ECC would generate
an additional $16$ bits to carry the $8$ parity bits, resulting in a total of
$88$ bus wires.

\subsection{Summary of Contributions}
\label{sec:intro_summary}

We propose an efficient method to perform joint CAC and ECC that uses fewer bus
wires than the state-of-the-art \cite{Sridhara05} and in addition can provide
better error correction. For the DDR4 example above, our proposed scheme for joint CAC and ECC uses
only an additional $10$ bits to carry the $8$ parity bits, resulting in a total
of $82$ bus wires. This constitutes a savings of $6$ bus wires over the state of
the art \cite{Sridhara05}.

The key ingredient in our approach for joint CAC and ECC is to identify what we
term \emph{free wires}. These are wires that can carry either a $0$ or a $1$
without violating the crosstalk-avoidance constraints and without influencing
what can be sent over adjacent bus wires. Suppose that there are $P$ such free
wires. Our joint CAC and ECC then operates as follows. We first encode the
information bits using a CAC to form $N-P$ encoded bits, skipping the free wires
during the encoding. We next generate $P$ parity bits for the $N-P$ CAC-encoded
bits using an ECC. We finally place the $P$ parity bits on the free wires
identified earlier. Notice that, by placing the parity bits on the free wires,
they automatically satisfy the CAC constraints, and hence we do not need to
duplicate or otherwise shield them in order to protect them against crosstalk as
done in \cite{Sridhara05}. In other words, the parities are embedded into the
bits to be transmitted. 

We point out that identification of the free wires is possible knowing only the
past bus state. Hence, the receiver can correctly determine which wires carry
the parity bits. Once the receiver knows the location of the parity bits, it can
decode the ECC and CAC. Here, we perform the ECC and CAC decoding jointly. This
joint decoding allows us to obtain better error correction than the approach
proposed in \cite{Sridhara05}, which decodes the ECC and CAC separately. In this
paper we use an irregular repeat-accumulate (IRA) code \cite{JKM00} as our
choice of ECC.  Since both the ECC and CAC employ local constraints we apply a
modified iterative belief-propagation decoder \cite{RiU08} to jointly decode the
transmitted codeword. 

In summary, the proposed method has two benefits over the state-of-the-art
\cite{Sridhara05}. First, our joint CAC-ECC encoding scheme at the transmitter,
by identifying free wires, uses fewer bus wires for transmission of data for the
same number of parity bits. Second, the joint CAC-ECC iterative decoding scheme at the
receiver allows for better error correction.

We provide a performance analysis for our proposed joint CAC-ECC scheme by
computing its rate, minimum distance, and iterative decoding threshold in
presence of random erasures. The factor graph \cite{Loe04} for our joint scheme
consists of both linear (ECC) and nonlinear (CAC) factor nodes. Hence, the
density evolution analysis is nonstandard and we believe this itself is an
interesting aspect of our work.

\subsection{Organization}
\label{sec:intro_organization}

The remainder of this paper is organized as follows.
Section~\ref{sec:background} provides background material on crosstalk-avoidance
coding. Section~\ref{sec:joint} introduces the joint embedded CAC-ECC code.
Sections~\ref{sec:analysis} and \ref{sec:density} analyze the proposed joint
embedded CAC-ECC code, evaluating its rate, minimum distance, and iterative
decoding thresholds.  The proofs of various statements are deferred to the appendices.

\section{Background on Crosstalk-Avoidance Codes}
\label{sec:background}

This section provides some background material on crosstalk-avoidance coding.
As mentioned earlier, due to capacitative coupling of adjacent bus wires,
certain bus transitions lead to longer transmission delays and higher energy
consumption than others \cite{SotiriadisThesis}. This effect is particularly
pronounced for opposing transitions on adjacent wires, i.e., a first wire
transitioning up (from $0$ to $1$) and an immediately adjacent wire
transitioning down (from $1$ to $0$).  Crosstalk-avoidance codes encode the data
to be transmitted over the bus to avoid these opposing transitions on adjacent
wires. This encoding reduces the transmission delay and energy by up to a factor
$1/2$ depending on the coupling coefficient at the cost of also reducing the
data rate \cite{SotiriadisThesis, SotiriadisChandrakasan02,
SotiriadisChandrakasan01}. 

The analysis of such codes was pioneered in~\cite{Victor01}. Consider a bus with
$N$ wires. Denote by $\bm{a} = (a_1, a_2, \dots ,a_N)\in\{0,1\}^N$ the past bus
state. We are interested in the number of sequences $\bm{b}\in\{0,1\}^N$
describing the next bus state that satisfy the crosstalk-avoidance constraint
with respect to $\bm{a}$. Clearly, the number of such sequences depends on the
past bus state $\bm{a}$. The best case is if $\bm{a}$ is either equal to all
ones or all zeros. Then, there can be no opposing transitions on adjacent wires
no matter what $\bm{b}$ is, and therefore the number of sequences $\bm{b}$
satisfying the CAC constraints is $2^N$. The worst case turns out to be if
$\bm{a}$ is an {\em alternating run} of 0 and 1, i.e., $0101\dots$ or
$1010\dots$, in which case the number of sequences $\bm{b}$ satisfying the CAC
constraints is $F(N+2)$, where $F(\cdot)$ denotes the Fibonacci numbers
\cite{Victor01}.

Alternating runs in $\bm{a}$ turn out to be crucial for the rate and decoder
analysis.  Consider a past bus state $\bm{a}$ consisting of two alternating
runs, e.g., $\dots01011010\dots$, and assume the first such alternating run has
length $d_1$ and the second one $d_2$. Note that the boundary of the two
alternating runs is either $11$ or $00$. Hence, there can be no opposing
transition regardless of the value of $\bm{b}$ between those two positions. More
precisely, if $\bm{b}_{1}$ and $\bm{b}_{2}$ denote any two vectors of $d_1$ and
$d_2$ bits, respectively, that satisfy the constraints imposed by the first
and second alternating runs, then the concatenation $\bm{b}$ of $\bm{b}_{1}$ and
$ \bm{b}_{2}$ is a vector of length $d_1+d_2$ that satisfies the constraints
imposed by the whole vector $\bm{a}$.  Thus, the boundary between alternating
runs decouples the problem of how to choose $\bm{b}$ into two noninteracting
subproblems of choosing the first $d_1$ bits and the second $d_2$ bits of $\bm{b}$.
Consequently, the number of possible $\bm{b}$ sequences is $F(d_1+2)F(d_2+2)$
\cite{Victor01}.

In general, assume $\bm{a}$ is (uniquely) parsed into maximal alternating runs
of lengths $\{d_m\}_{m=1}^M$. Then the number of possible $\bm{b}$ sequences is
$\prod_{m=1}^M F(d_m+2)$ \cite{Victor01}. The rate $\Rcac(\bm{a})$ of the CAC
code with past bus state $\bm{a}$ is thus given by
\begin{equation}
    \label{eq:rcac}
    \Rcac(\bm{a}) = \frac{1}{N}\sum_{m=1}^M \log F(d_m+2).
\end{equation}

For past bus state $\bm{a}$ generated uniformly at random over $\{0,1\}^N$, it
is not hard to show that the expected number of alternating runs of degree $d$
is $N2^{-d-1}(1+o(1))$ asymptotically as $N\to\infty$ and that the actual number of
alternating runs concentrates around this distribution. Thus, for this random
choice of $\bm{a}$, the CAC has asymptotic rate
\begin{equation*}
    \Rcac(\bm{a}) = \sum_{d=1}^{\infty}2^{-d-1}\log F(d+2) \approx 0.824
\end{equation*}
with high probability as $N\to\infty$.

\section{Embedded Joint CAC-ECC Coding Scheme}
\label{sec:joint}

In this section we introduce our embedded joint CAC and ECC scheme. One of the
key concepts for this scheme are what we term free wires, which are introduced
in Section~\ref{sec:joint_free}. The encoder and decoder operation are described
in Sections~\ref{sec:joint_encoder} and \ref{sec:joint_decoder}.

\subsection{Free Wires}
\label{sec:joint_free}

Assume as before that the bus has $N$ wires and that the past bus state is
$\bm{a} = (a_1, a_2, \dots ,a_N)$.  We want to set the bus to its next state,
denoted again by $\bm{b} = (b_1, b_2, \dots, b_N)$.  A wire $n$ such that
 all the three wires, $a_{n-1}$, $a_n$ and $a_{n+1}$ have the same value is called a {\em free} wire.
For such a free wire, we can set the next value $b_n$ to be either $0$ or $1$
without violating the crosstalk constraints and without affecting the values
$b_{n-1}$, $b_{n+1}$ that can be sent over the two adjacent wires. 

\begin{figure}[htbp]
    \centering
    \includegraphics{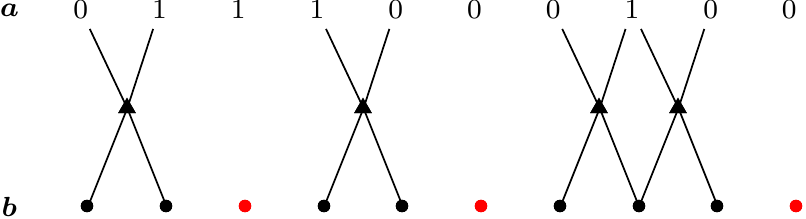}
    \caption{Factor graph depicting free wires (drawn in red). $\bm{a}$ and
    $\bm{b}$ are the past and future bus state, respectively.}
    \label{fig:free}
\end{figure} 

The identification of the free wires is best understood using a factor graph \cite{Loe04} as
shown in Fig.~\ref{fig:free}. On the top in the figure we show the past bus
state $\bm{a}$, and on the bottom we show the next bus state $\bm{b}$. The wires
in the next bus state $\bm{b}$ are denoted by filled circles.  The triangles
in the middle represent the crosstalk-avoidance constraints imposed by the past
bus state $\bm{a}$.  For example, $(a_1, a_2) = (0, 1)$ requires that
$(b_1,b_2) \neq (1,0)$ since otherwise we would have an opposing transition on
these two adjacent wires.

Notice that there are some values of $\bm{b}$ that are not connected to any
constraint.  In Fig.~\ref{fig:free}, the wires $b_3$, $b_6$, $b_{10}$ (indicated
by red filled circles) do not have any constraints associated with them.  These
are the free wires. Thus, the wires $b_3$, $b_6$, $b_{10}$ can be set to
any value without affecting adjacent wires and without violating any
crosstalk constraints.  Note that the crosstalk constraints are local:
each constraint is connected to two wires of the past bus state and
two wires of the next bus state. We also emphasize that the crosstalk
constraints are nonlinear, since only one out of four possible values of two
adjacent wires is prohibited.

\subsection{Encoder}
\label{sec:joint_encoder}

\begin{figure}[htbp]
    \centering
    \includegraphics{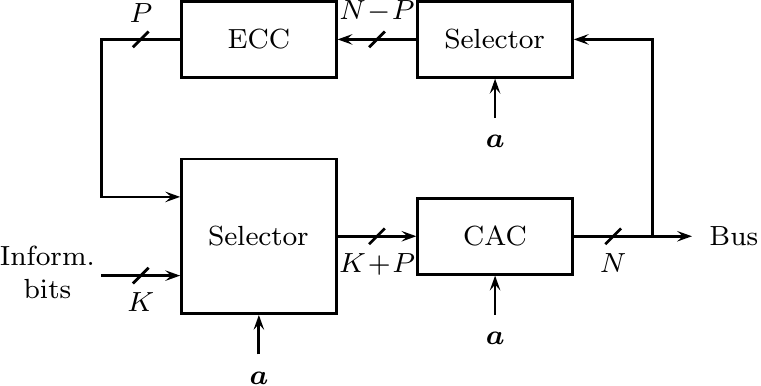}
    \caption{Embedded joint CAC-ECC encoder proposed in this paper.}
    \label{fig:encoder}
\end{figure} 

The encoder architecture is shown in Fig.~\ref{fig:encoder}.  Given the past
state $\bm{a}$, the encoder first determines the positions of the free wires.
The information bits are passed through the CAC and placed on the non-free
wires. The ECC, assumed to be a linear code here and throughout, then computes
the parity bits of the data placed on the non-free wires and places them on the
free wires. 

More formally, let us assume for the moment that the number of free wires is
exactly equal to the desired number of parity bits. Then, as shown in the
Fig.~\ref{fig:encoder}, the encoder first identifies the $P$ free wires. The $K$
information bits are encoded using a CAC and are placed on the $N-P$ non-free
wires. The $N-P$ CAC-encoded bits are then fed to the ECC, which generates $P$
parity bits. These parities are then placed on the $P$ free wires.  The combined
$N-P$ CAC-encoded bits and the generated $P$ parities are sent across the bus.

If the number of free wires is more than the number of parity bits, then the
selector uses some of the free wires to carry CAC-encoded information bits. If
the number of free wires is less than the number of parity bits, then parity
bits are forced on the non-free wires by appropriately shielding
them.\footnote{We can use two wires to send one parity bit. We set the first
wire to the bit that was sent in the same wire in the past state, and we
send the parity bit over the second wire. This strategy of shielding always
satisfies the opposing-transition constraint. It is this type of shielding,
using two adjacent wires to carry a single parity, that is used in the joint
encoding scheme \cite{Sridhara05} for \emph{all} parities.}

In practice, the channel typically operates at very high SNR, and hence we can
restrict ourselves to high-rate ECCs, so that the number of parity bits is
small. We will show later that when the past bus state $\bm{a}$ is generated
uniformly at random from $\{0,1\}^N$, then the number of free wires is typically
much larger than the desired number of parity bits. Thus, the shielding
technique would be rarely used, and our scheme will be efficient most of the
time.

So far, we have made no assumptions about the ECC being used except for
linearity. To keep decoding complexity manageable, we would like to use a
low-density  ECC. However, a standard low-density parity-check (LDPC) code cannot
be directly combined with the architecture described here since the resulting
code will not have vanishing probability of block error. Instead, we propose to
use an IRA code \cite{JKM00}, in which the parities are further accumulated. 

\begin{figure}[htbp]
    \centering
    \includegraphics{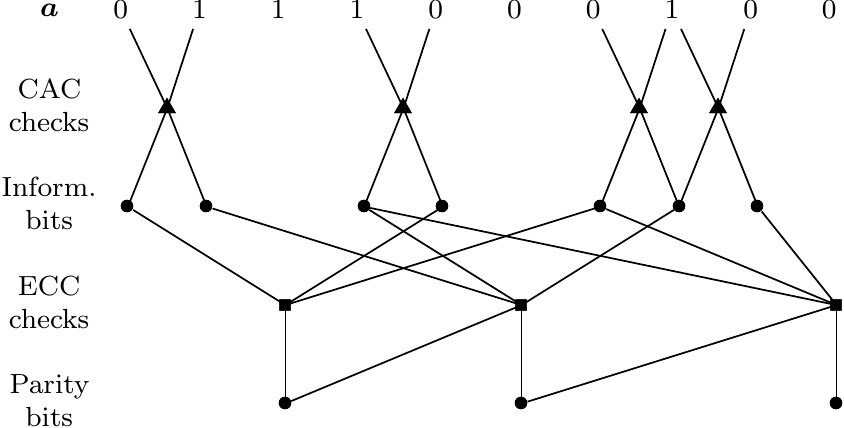}
    \caption{Factor graph for joint embedded CAC and ECC. For clarity, the
        parity bits are shown separately from the (CAC encoded) information bits.
        Both of them together represent the $\bm{b}$ vector of bits to be
        transmitted over the bus.}
    \label{fig:factor2}
\end{figure} 

An example of the factor graph for this construction is shown in
Fig.~\ref{fig:factor2}.  In this figure, the parity-check constraints of the ECC
are shown by black squares. An edge between a parity-check constraint and a wire
in the next bus state denotes the participation of the bit (on that wire) in
that parity-check constraint. The edges associated to the past bus state
$\bm{a}$ are half-edges, describing the (fixed and known) bits in the past bus state.

\subsection{Decoder}
\label{sec:joint_decoder}

To explain the receiver operation we assume that the past bus state $\bm{a}$ is
known correctly at the receiver. This allows the receiver to determine the
position and number of free wires. We explain only the case when the number of
free wires is exactly equal to the number of parity bits. The other cases follow
in a similar fashion.  The receiver performs a joint iterative decoding of the
CAC and ECC to reliably recover the information bits.

Consider the joint factor graph shown in Fig.~\ref{fig:factor2}.  The goal of
the decoder is to determine the information bits in $\bm{b}$ given the
noise-corrupted bus output.  We will use a modified belief-propagation decoder,
which passes messages along the joint factor graph. This modification is needed
to take the nonlinear CAC constraints into account. Joint decoding of the
CAC and ECC has the advantage that it provides additional error correction
compared to the standard approach of only using the ECC for error correction.
Indeed, the CAC part of the factor graph provides extrinsic information that can
be used by the message-passing decoder.

The $N$ bits in $\bm{b}$, encoded as mentioned above, are transmitted over the
bus and face random erasures. The input of the decoder is an $N$-tuple with each
entry in $\{0,1,?\}$, where we use $?$ to denote erasures, describing the output
of the bus. The decoder is best explained in the language of the joint factor
graph consisting of various nodes (cf. Fig.~\ref{fig:factor2}). We refer to the
wires in the bus, over which the CAC encoded information bits are sent, as the
information variable nodes of the factor graph. The parity-bits of the IRA are 
called parity variable nodes.  The CAC and ECC constraints are called
CAC-check and ECC-check nodes, respectively.  We consider the following message
updates at the different nodes in the factor graph.

\begin{itemize}
    \item{{\em Information and parity variable-node updates:}} Send an erasure message
        on an outgoing edge if all the incoming messages (including the channel
        value) are erasures, otherwise send the value (either $0$ or $1$) of the
        non-erased incoming message.

    \item{{\em CAC node update:}} Each CAC-constraint node has degree two (cf.
        Fig.~\ref{fig:factor2}) and has two bits of the past bus state as
        other inputs through half-edges. We distinguish the two (full) edges,
        connecting the current bus state, by calling them {\em left edge} and {\em
        right edge}, respectively.  Let us assume for the moment that the past
        bus state for that constraint node is $(0,1)$ (see the first CAC
        constraint in Fig.~\ref{fig:factor2}). If the incoming message on the
        left edge is $1$, then the outgoing message on the right edge is set to
        $1$ because the CAC constraint prohibits the value $(1,0)$ in the
        current bus state. Otherwise, the outgoing message on the right edge is
        set to $?$. If the incoming message on the left edge is $0$, then the
        outgoing message on the right edge is set to $0$ again because the CAC
        constraint prohibits the value $(1,0)$ in the current bus state.
        Otherwise the outgoing message on the left edge is set to $?$. The
        CAC-update rule is analogous for past bus state $(1,0)$.

    \item{{\em ECC check node update:}} An outgoing message is a non-erasure
        (either $0$ or $1$) only if all the incoming messages are non-erased, in
        which case the outgoing message is set to be the XOR of the incoming
        messages. Otherwise, the outgoing message is set to $?$.
\end{itemize}

We use the following decoding schedule:
\begin{enumerate}
    \item Send messages from the information variable nodes to the CAC check
        nodes.

    \item Send messages from the CAC check nodes to the information
        variable nodes.

    \item Send messages from the information variable nodes to the ECC check
        nodes. 

    \item Send messages between the ECC check nodes and the parity variable
        nodes until the messages converge on the accumulate chain. 

    \item Send messages from the ECC check nodes to the information variable
        nodes.
\end{enumerate}

This process is repeated until all messages on all edges converge.  Initially,
the messages sent from the variable nodes to the CAC check nodes are the channel
observations. The final value of an information variable node is declared to be
an erasure if the incoming messages on all the edges connected to that node are
in erasure, otherwise it is set to the value of any non-erased incoming message.

\section{Rate and Minimum-Distance Analysis}
\label{sec:analysis}

We begin with the computation of the rate of our embedded joint CAC and ECC
scheme. We will compare this with the rate for the shielded joint CAC and ECC
from~\cite{Sridhara05}. We next analyze the minimum distance of the proposed
embedded joint CAC and ECC scheme. All the results in this section pertain to
arbitrary linear ECCs (not necessarily IRA codes).

\subsection{Rate}
\label{sec:analysis_rate}

Recall from~\eqref{eq:rcac} that $\Rcac(\bm{a})$ denotes the rate of the single
CAC for past bus state $\bm{a}$. We also denote by $\Recc$ the rate of the
single ECC. The next theorem provides the rate of the shielded CAC-ECC
encoder from~\cite{Sridhara05}.

\begin{theorem}
    \label{thm:rate_shielded}
    The rate $\Rs(\bm{a})$ of the shielded joint CAC-ECC encoder
    from~\cite{Sridhara05} is 
    \begin{equation*}
        \Rs(\bm{a}) = \Rcac(\bm{a})\bigl(2\Recc^{-1}-1\bigr)^{-1}.
    \end{equation*}
\end{theorem}

The rate for the proposed  embedded joint CAC and ECC scheme is given by the
following theorem.

\begin{theorem}
    \label{thm:rate_embedded}
    Assume that the past bus state $\bm{a}$ has sufficient number of free wires
    to carry all ECC parities. Then the rate $\RE(\bm{a})$ of the embedded joint
    CAC-ECC encoder proposed in this paper is
    \begin{equation*}
        \RE(\bm{a}) = \Rcac(\bm{a})+\Recc-1.
    \end{equation*}
\end{theorem}

The proofs of Theorems~\ref{thm:rate_shielded} and~\ref{thm:rate_embedded} are
provided in Appendices~\ref{sec:proofs_shielded} and~\ref{sec:proofs_embedded}.
We can now compare the two rate expressions
\begin{align*}
    \Rs(\bm{a}) & = \Rcac(\bm{a})\bigl(2\Recc^{-1}-1\bigr)^{-1}, \\
    \RE(\bm{a}) & = \Rcac(\bm{a})+\Recc-1,
\end{align*}
for the shielded and the embedded joint CAC-ECC schemes. We start with a
numerical example from the DDR4 SDRAM standard~\cite{DDR4}.

\begin{example}
    \label{eg:ddr4}
    The DDR4 interconnect communication standard uses a rate $\Recc = 0.9$
    cyclic-redundancy check ECC. Assuming the past bus state $\bm{a}$ is
    generated uniformly over $\{0,1\}^N$, the CAC rate is approximately
    $\Rcac(\bm{a}) \approx 0.824$ with high probability for large enough $N$ (cf. Section~\ref{sec:background}). Hence,
    \begin{align*}
        \Rs(\bm{a}) & \approx 0.674, \\
        \RE(\bm{a}) & \approx 0.724.
    \end{align*}
    Assume we want to transmit $59$ data bits (resulting in
    $\ceil{59/\Rcac(\bm{a})} = 72$ CAC encoded bits as mentioned in
    Section~\ref{sec:intro}).  The shielded joint CAC-ECC encoder
    from~\cite{Sridhara05} requires $\ceil{59/\Rs(\bm{a})} = 88$ wires. The
    embedded joint CAC-ECC encoder proposed here requires only
    $\ceil{59/\RE(\bm{a})} = 82$ wires, thereby saving $6$ wires for the same
    number of parity bits conveyed.
\end{example}

As can be seen from this example, the regime of practical interest is when the
error correction code has rate $\Recc$ close to one. Assume then that $\Recc =
1-\delta$ for some small $\delta > 0$. Then,
\begin{align*}
    \Rs(\bm{a}) 
    & = \Rcac(\bm{a})\frac{1-\delta}{1+\delta} \\
    & \approx \Rcac(\bm{a})(1-2\delta) \\
    & = \Rcac(\bm{a})-2\Rcac(\bm{a})\delta. \\
    \shortintertext{On the other hand,} \\
    \RE(\bm{a}) & = \Rcac(\bm{a})-\delta.
\end{align*}

For a concrete example, assume again that $\bm{a}$ is generated uniformly over
$\{0,1\}^N$ so that $\Rcac(\bm{a}) \approx 0.824$ with high probability for
large enough $N$. Then the embedded approach proposed here outperforms the
shielded approach from~\cite{Sridhara05} by approximately $0.648\delta$ (as long
as it has sufficient free wires to hold the parities for the embedding which is
the case with high probability when $\delta < 1/4$).

More generally, it can be shown that $\RE(\bm{a}) \geq \Rs(\bm{a})$ for any past
bus state $\bm{a}$ with sufficient number of free wires. Thus, the proposed
embedded joint CAC-ECC scheme always outperforms the shielded joint CAC-ECC
scheme from~\cite{Sridhara05}. The proof of this fact is reported in
Appendix~\ref{sec:proofs_comparison}.

\subsection{Minimum Distance}
\label{sec:analysis_dmin}

Denote by $\decc$ the minimum distance of the ECC.  The next theorem shows that
the minimum distance of the embedded joint CAC-ECC code is the same as that of
the ECC alone. 

\begin{theorem}
    \label{thm:dmin}
    Assume that the past bus state $\bm{a}$ has sufficient number of free wires
    to carry all ECC parities. Then the minimum distance $\de(\bm{a})$ of the embedded
    joint CAC-ECC encoder  is
    \begin{equation*}
        \de(\bm{a}) = \decc.
    \end{equation*}
\end{theorem}

The proof of Theorem~\ref{thm:dmin} is reported in
Appendix~\ref{sec:proofs_dmin}. The theorem shows that joint CAC-ECC encoding
does not increase minimum distance. In other words, the additional redundancy
introduced by the CAC requirement does not translate into increased minimum
distance. This is, of course, somewhat disappointing. However, as we will see in
the next section, the CAC redundancy is nonetheless beneficial for error
correction, since it \emph{does} increase the iterative decoding threshold.

\section{Iterative Decoding Threshold}
\label{sec:density}

In this section we will assume that the ECC is an IRA code as described in
Section~\ref{sec:joint_encoder}. We perform a density evolution analysis
\cite{RiU08, RiU01, RSU01} of the iterative decoder described in
Section~\ref{sec:joint_decoder}. This analysis allows us to predict the
performance of our embedded encoding scheme in the presence of random erasures.
We begin in Section~\ref{sec:density_notation} by formally introducing the
ensembles over which the analysis is performed. The casual reader may wish to
skip this section and proceed directly to Section~\ref{sec:DEanalysis}, which
reports the density evolution analysis.

\subsection{Ensembles}
\label{sec:density_notation}

Density evolution is an average analysis and predicts the probability of error
when averaged over an ensemble of codes. We next describe the ensemble of codes
with the help of Fig.~\ref{fig:factor2} in Section~\ref{sec:joint_encoder}.

\subsubsection{Ensemble of Past Bus States}

We start with the probabilistic description of the past bus state $\bm{a}$. A
natural assumption would be that $\bm{a}$ is generated uniformly at random from
$\{0,1\}^N$. As we have seen in Section~\ref{sec:background}, with this
assumption the expected number of alternating runs in $\bm{a}$ of length $d$ is
asymptotically $N2^{-d-1}$. Unfortunately, this assumption results in a
cumbersome analysis, and we instead adopt a modified assumption on the
generation of $\bm{a}$ that is easier to handle and asymptotically equivalent as
$N\to\infty$.

Recall that the past state $\bm{a}$ can be thought of as a collection of
alternating runs of various lengths, with free wires being alternating runs of
length one. In our modified assumption, we will generate $\bm{a}$ as the
concatenation of independently generated such alternating runs of random length.

Fix the rate for the ECC to be $\Recc\in (3/4,1]$.  Generate $\bm{a}$ as two
different parts $\bm{a}_1$ and $\bm{a}_2$. The first part consists of
$N(\Recc-1/2)$ independently generated alternating runs. Each such alternating
run has random length $D$ with identical probability given by
\begin{equation*}
    \Pp(D = d) \defeq
    \begin{cases}
        \frac{2\Recc-3/2}{2\Recc-1}, & \text{for $d = 1$},\\
        \frac{2^{-d}}{2\Recc-1}, & \text{for $d \in \{2,3,\dots\}$}.
    \end{cases}
\end{equation*}

Let $\ell(\bm{a}_1)$ be the length of this first part. The expected value of
this length is
\begin{equation*}
    N(\Recc-1/2)\sum_{d=1}^\infty d\Pp(D=d) 
    = \frac{N(\Recc-1/2)}{2\Recc-1} \biggl(2\Recc-3/2+\sum_{d=2}^\infty d2^{-d}\biggr)
    = N\Recc.
\end{equation*} 
The bits in $\bm{b}$ corresponding to $\bm{a}_1$ will be used to carry the CAC
encoded information bits.

The second part $\bm{a}_2$ is generated as $\ell(\bm{a}_1)(1-\Recc)/\Recc$ alternating
runs of length one. Thus, the length $\ell(\bm{a}_2)$ of this second part is
\begin{equation}
    \label{eq:la2}
    \ell(\bm{a}_2) = \ell(\bm{a}_1)(1-\Recc)/\Recc
\end{equation}
and has expected value
\begin{equation*}
    N\Recc(1-\Recc)/\Recc = N(1-\Recc).
\end{equation*} 
The bits in $\bm{b}$ corresponding to $\bm{a}_2$ will be used to carry the
accumulated parity bits. 

We generate the complete past bus state $\bm{a}$ by randomly interleaving the
alternating runs in $\bm{a}_1$ and in $\bm{a}_2$, i.e., with uniform
distribution over all possible interleavings of the corresponding runs.
The past state $\bm{a}$ has thus expected length
\begin{equation*}
    N\Recc+N(1-\Recc) = N
\end{equation*}
as required.  

We also point out that the expected number of alternating runs in
$\bm{a}$ of degree $d\geq 2$ is 
\begin{equation*}
    N(\Recc-1/2)\frac{2^{-d}}{2\Recc-1} = N2^{-d-1},
\end{equation*}
and of degree $d = 1$ is
\begin{equation*}
    N(\Recc-1/2)\frac{2\Recc-3/2}{2\Recc-1}+N(1-\Recc) = N2^{-2}.
\end{equation*}
Thus, asymptotically as $N\to\infty$, we expect that around $N2^{-d-1}$ of
alternating runs in $\bm{a}$ have length $d$ for any $d\in\{1,2,\dots\}$.
Thus, the natural way of generating the past bus state as being chosen uniformly
at random from $\{0,1\}^N$ has the same asymptotic distribution of alternating
runs. This also shows that, if $\Recc > 3/4$, then there are asymptotically
sufficient number of free wires to hold all parities.
 
Note that so far we have only described the length of the alternating runs of
$\bm{a}$, but not their actual values. These are chosen by selecting the first
bit of $\bm{a}$ uniformly over $\{0,1\}$ independently of everything else. The
remaining bits of $\bm{a}$ are then fully specified by the alternating run
structure.

\subsubsection{Ensemble of Joint CAC and ECC Codes}

We consider an ECC that is randomly selected from an ensemble of IRA codes as
described next. The information bits are first encoded using the CAC on the
wires in $\bm{b}$ corresponding to the past bus state $\bm{a}_1$. For the
purpose of analysis, we assume that the CAC-encoded information bits are chosen
uniformly at random from the collection of all CAC codewords $\Ccac(\bm{a}_1)$
compatible with $\bm{a}_1$. These bits are then fed to the IRA code as
systematic bits. Then, the free wires in $\bm{b}$ corresponding to the past bus
state $\bm{a}_2$ are set to the parity bits of the IRA using the code's
parity-accumulate structure.

We start with the description of the connection of CAC check nodes and
information variable nodes. Unlike what is depicted in Fig.~\ref{fig:factor2},
we represent each alternating run in $\bm{a}_1$ by a single CAC check node of
degree $d$.  This CAC check node is connected to the corresponding $d$
information variable nodes in $\bm{b}$. 

We continue with the description of the accumulate structure of the IRA.  We
generate $\ell(\bm{a}_2)$ ECC checks, one for each of the parity variable nodes.  We
pair the ECC check nodes and the parity variable nodes by connecting them with an
edge. Note that the number of ECC checks is a random quantity here, chosen to
match the random realization of $\ell(\bm{a}_2)$. We further add an accumulation
structure on the parity bits as shown in Fig.~\ref{fig:factor2}.  This
accumulation structure is needed to ensure that the code has a nontrivial
threshold. 

We next describe the construction of the LDPC code linking the $\ell(\bm{a}_1)$
information variable nodes with the $\ell(\bm{a}_2)$ ECC check nodes.  The
design rate of this LDPC code is
\begin{align}
    \label{eq:rldpc}
    \Rldpc 
    & = 1-\frac{\ell(\bm{a}_2)}{\ell(\bm{a}_1)} \notag\\
    & \approx 1-\frac{1-\Recc}{\Recc} \notag\\
    & = 2-\frac{1}{\Recc},
\end{align}
where the second equality follows from~\eqref{eq:la2} and where the
approximation is due to the concentration of $\ell(\bm{a}_1)$ and
$\ell(\bm{a}_2)$ around their expected values as $N\to\infty$.  Note that, since
$\Recc\in(3/4, 1]$, this design rate satisfies $\Rldpc\in (2/3, 1]$. 

Denote by $\lambda(x)$ and $\rho(x)$ the edge-perspective degree distributions
of the LDPC code.  Denote by $(L(x), R(x))$ the normalized variable and check
degree distribution generator polynomials from a node perspective corresponding
to the normalized degree distribution generator polynomials from an edge
perspective $(\lambda(x),\rho(x))$. The degree distributions are chosen to
satisfy the desired design rate,
\begin{equation*}
    2-\frac{1}{\Recc} = \Rldpc = 1 - \frac{L'(1)}{R'(1)}
\end{equation*}
or, more succinctly,
\begin{equation*}
    \frac{L'(1)}{R'(1)} = \frac{1}{\Recc}-1.
\end{equation*}

The LDPC code linking the ECC checks with the information variable nodes is now
generated by randomly connecting the information variable nodes to the ECC check
nodes as in the standard LDPC ensemble using the degree distributions $(L(x),
R(x))$.

\subsection{Density Evolution Analysis}
\label{sec:DEanalysis}

Recall that $\lambda(x), \rho(x)$ denote the edge-perspective degree
distributions of the irregular LDPC part of the IRA ensemble, which describe the
connections between the information variable nodes and the ECC check nodes.
Further, let $\tilde{\rho}_{d}$ denote the edge-perspective degree distribution
of the CAC part. More precisely, $\tilde{\rho}_{d}$ is the probability that a
randomly picked edge (in the CAC part) is connected to a CAC check node of
degree $d$.  From the previous section, the degree distribution $\tilde{\rho}_d$
can be evaluated as
\begin{equation*}
    \tilde{\rho}_d = 
    \begin{cases}
        1-3/(4\Recc), & \text{for $d=1$}, \\
        d2^{-d-1}/\Recc, & \text{for $d\in\{2,3\dots\}$}.
    \end{cases}
\end{equation*}

For $d\in\{1,2,\dots\}$, denote by $F(d)$ the Fibonacci numbers as before.  For
each $d\in\{2,3,\dots\}$, $i\in\{1,2,\dots,d\}$ define the polynomial 
\begin{equation*}
    p_{d,i}(x) \defeq (1-x)p_{d,i}^{(1)}+(1-x^2)p_{d,i}^{(2)},
\end{equation*}
with coefficients
\begin{align*}
    p_{d,1}^{(1)} & \defeq p_{d,d}^{(1)} \defeq \frac{F(d-1)}{F(d+2)}, \\
    p_{d,1}^{(2)} & \defeq p_{d,d}^{(2)} \defeq 0, \\
    p_{d,i}^{(1)} & \defeq \frac{F(i-1)F(d-i+1)+F(i)F(d-i)}{F(d+2)}, \quad\text{for $i\in\{2,3,\dots,d-1\}$}, \\
    p_{d,i}^{(2)} & \defeq \frac{F(i-1)F(d-i)}{F(d+2)}, \quad\text{for $i\in\{2,3,\dots,d-1\}$}.
\end{align*}

\begin{figure}[htbp]
    \centering
    \includegraphics{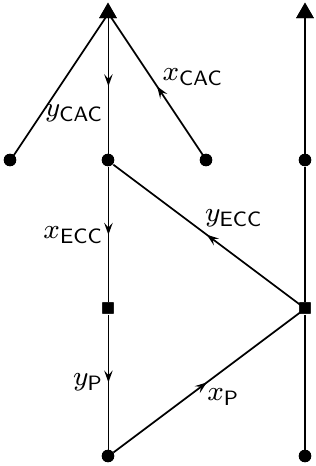}
    \caption{Erasure probabilities as tracked by the density evolution.}
    \label{fig:messages}
\end{figure} 

Consider now the messages being passed on the factor-graph edges by the decoder.
Let $\xecc, \yecc$, $\xcac, \ycac$, $\xp, \yp$ denote the fraction of 
erasure messages on the various edges as indicated in Fig.~\ref{fig:messages}.
The next theorem provides the density evolution equations for these erasure
probabilities.

\begin{theorem}
    \label{thm:density}
    The density-evolution equations for the iterative erasure decoder are
    \begin{align*}
        \xecc^{+} & = \varepsilon \ycac^{-}\lambda(\yecc^{-}), \\
        \yecc^{+} & = 1-(1-\xp^{-})^2\rho(1-\xecc^{-}), \\
        \xp^{+} & = \varepsilon \yp^{-}, \\
        \yp^{+} & = 1-(1-\xp^{-})R(1-\xecc^{-}), \\
        \xcac^{+} & = \varepsilon L(\yecc^{-}), \\
        \ycac^{+} & = 1-\sum_{d=2}^\infty \frac{\tilde{\rho}_{d}}{d}\sum_{i=1}^d
        p_{d,i}(\xcac^{-}).
    \end{align*}
    Here the $+$ and $-$ superscripts indicate outgoing and incoming messages,
    respectively.
\end{theorem}

The proof of Theorem~\ref{thm:density} is reported in
Appendix~\ref{sec:proofs_density}. Recall that for each decoding iteration the
accumulation part of the decoder is run until convergence.  Therefore, $\xp$ in
Theorem~\ref{thm:density} satisfies the following fixed-point equation in every
iteration:
\begin{equation*}
    \xp  = \varepsilon\bigl(1-(1-\xp)R(1-\xecc)\bigr).
\end{equation*}
This can be solved for $\xp$ as
\begin{equation*}
    \xp  = \frac{\varepsilon\bigl(1-R(1-\xecc)\bigr)}{1-\varepsilon R(1-\xecc)}.
\end{equation*}
By substituting this back into the system of equations in
Theorem~\ref{thm:density}, it is not hard to see that we obtain a one-dimensional density evolution curve
tracking solely the parameter $\xecc$. 

\begin{example}
    \label{eg:de}
    Choose the LDPC code as a regular $(3,12)$ code. The design rate of
    this code is $\Rldpc = 1-3/12 = 3/4$. Together with the accumulated
    parities, the error correcting part of the embedded joint CAC-ECC scheme has
    rate
    \begin{equation*}
        \Recc = \frac{1}{2-\Rldpc} = 0.8
    \end{equation*}
    by \eqref{eq:rldpc}. Applying Theorem~\ref{thm:rate_embedded} and using that
    $\Rcac(\bm{a}) \approx 0.824$ as seen in Section~\ref{sec:background}, the
    embedded joint CAC-ECC scheme has rate equal to
    \begin{equation*}
        \RE(\bm{a}) 
        = \Rcac(\bm{a})+\Recc-1
       \approx 0.624
    \end{equation*}
    with high probability for large block length $N$.

    \begin{figure}[htbp]
        \centering
        \includegraphics{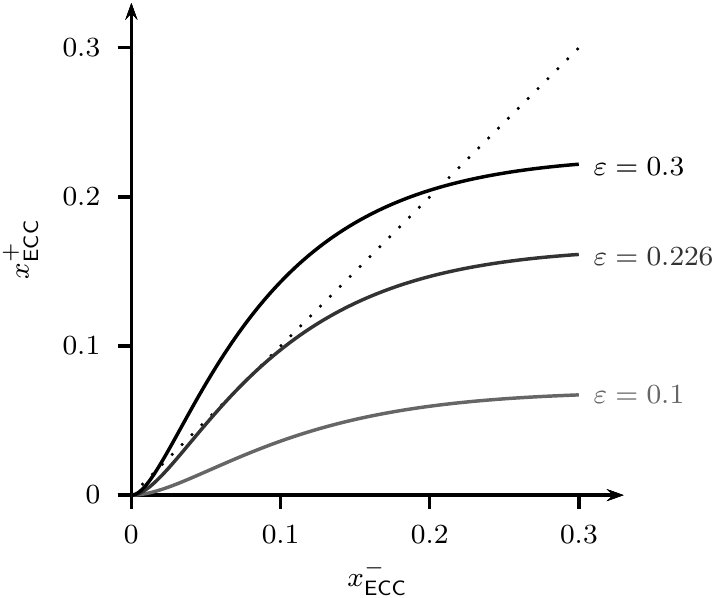}
        \caption{Density evolution for joint CAC-ECC decoding using a regular
        $(3,12)$ LDPC code.}
        \label{fig:de2}
    \end{figure} 

    Fig.~\ref{fig:de2} shows the density evolution for this setting. The
    threshold for the erasure channel is around $0.226$. Observe that, for an optimal code and MAP
    decoding, the threshold for the ECC part \emph{alone} is $1-\Recc = 0.2$.
    Hence, even with suboptimal codes and decoding as performed here, the
    \emph{joint} decoding of the ECC and CAC allows us to operate above this
    threshold. Thus, while the minimum distance of the joint CAC-ECC scheme is
    the same as the minimum distance of the ECC alone by Theorem~\ref{thm:dmin},
    joint CAC-ECC decoding nonetheless increases the decoding threshold thanks to the valuable
extrinsic information provided by the CAC.

    \begin{figure}[htbp]
        \centering
        \subfigure[Probability of bit error $P_b$.]{\includegraphics{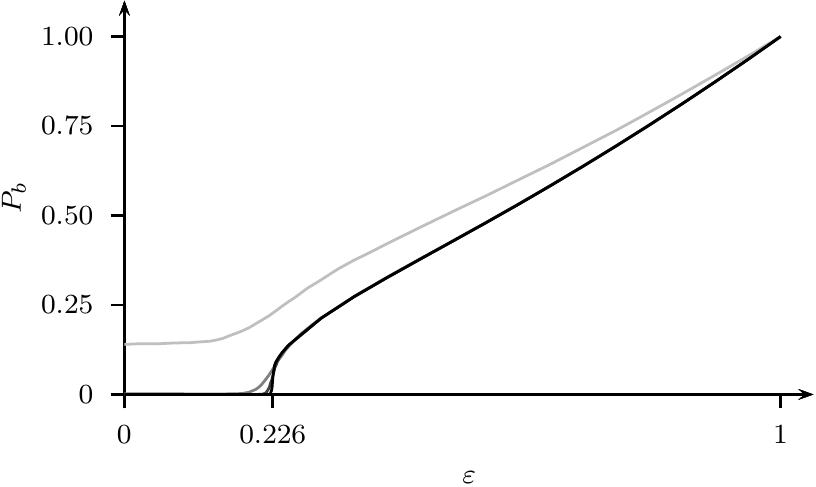}}
        \subfigure[Probability of block error $P_e$.]{\includegraphics{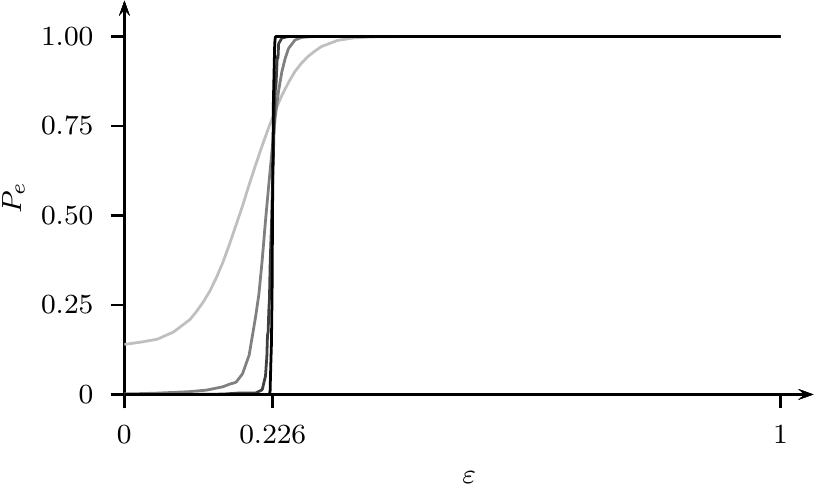}}

        \caption{Monte-Carlo simulations of the probabilities of bit and block
            error for a regular $(3,12)$ LDPC code. The plots are for block
            lengths of $N\in\{10^2, 10^3, 10^4, 10^5\}$ (from light gray to
            black). A phase transition at the analytically derived
            threshold of $0.226$ is apparent.}

        \label{fig:perr}
    \end{figure} 

    We complement the asymptotic analysis with simulation results for a joint
    CAC+ECC encoder and decoder. In our simulation setting, we use a fixed block
    length of $N$ and generate the past bus state $\bm{a}$ uniformly at random
    over $\{0,1\}^N$. Since we are interested in the asymptotic behavior,
    we do not handle the case with insufficient free wires and simply declare an
    error in this case. Fig.~\ref{fig:perr} shows the resulting probabilities of
    bit and block error for a regular $(3,12)$ LDPC code for $N\in\{10^2, 10^3,
    10^4, 10^5\}$. Both error probabilities exhibit a clear phase transition at
    the theoretically derived threshold of $0.226$ as expected. These simulation
    results also validate the modified ensemble (with variable length of the
    prior bus state $\bm{a}$) used for the asymptotic analysis.

    It is worth commenting on the behavior of the curves around erasure rate
    $\varepsilon = 0$. The error probability is not zero in this regime due to
    insufficient free wires being treated as error events. For $N=10^2$ and
    $\Recc = 0.8$ as used here, a Binomial approximation estimates this error
    event as happening with probability $0.149$. The simulation results for
    $N=10^2$ yield $P_e(0) \approx 0.141$. As the block length increases, this
    error event has exponentially decreasing probability and is no longer
    apparent on the plots.
\end{example}

\begin{appendices}

\section{Proof of Theorem~\ref{thm:rate_shielded}}
\label{sec:proofs_shielded}

\begin{figure}[htbp]
    \centering
    \includegraphics{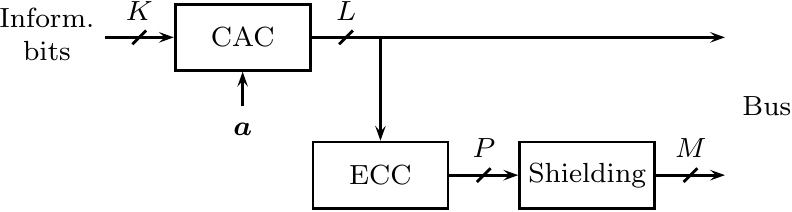}
    \caption{Shielded joint CAC-ECC encoder from~\cite{Sridhara05}.}
    \label{fig:shielded}
\end{figure} 
The shielded joint CAC-ECC encoder from~\cite{Sridhara05} has the structure
shown in Fig.~\ref{fig:shielded}. By definition,
\begin{align*}
    \Rs & = \frac{K}{L+M}, \\
    \Rcac & = \frac{K}{L}, \\
    \Recc & = \frac{L}{L+P}.
\end{align*}
Since the parity bits are duplicated we have,
\begin{align*}
 \frac{P}{M} & =    \frac{1}{2}.
\end{align*}
From the equation for $\Recc$ we can solve for $P/L$ as 
\begin{equation*}
    \frac{P}{L} = \Recc^{-1}-1.
\end{equation*}
Now,
\begin{align*}
    \Rs^{-1}
    & = \frac{L}{K}+\frac{M}{K} \\
    & = \frac{L}{K}+\frac{L}{K}\cdot\frac{P}{L}\cdot\frac{M}{P} \\
    & = \Rcac^{-1}+\Rcac^{-1}\cdot\bigl(\Recc^{-1}-1\bigr)\cdot2 \\
    & = \Rcac^{-1}\bigl(2\Recc^{-1}-1\bigr),
\end{align*}
so that 
\begin{equation*}
    \Rs = \Rcac\bigl(2\Recc^{-1}-1\bigr)^{-1}
\end{equation*}
as claimed.\hfill\IEEEQED

\section{Proof of Theorem~\ref{thm:rate_embedded}}
\label{sec:proofs_embedded}

Since we send $K$ information bits over the $N$-wire bus, the rate of the joint
CAC and ECC scheme is given by $\RE  = K/N$. We want to express this in terms of
the rate of the CAC and the ECC. Since $N-P$ bits on the non-free wires are used
to generate the $P$ parity bits, and since we assume that $\bm{a}$ has a
sufficient number of free wires to carry those parities, we have
\begin{align*}
\Recc & = \frac{N-P}{N}.
\end{align*}

We next compute the rate of the CAC encoder. Recall that this rate
depends on the past state $\bm{a}$. Since the past bus state 
gives rise to $P$ free wires, the CAC encoder can put any information on these
$P$ wires without affecting the setting of bits on the non-free wires. Thus, we
have $K+P$ information bits\footnote{Although we put parity information on the
free wires, the CAC encoder can, a priori, allow $2^K2^P$ $N$-tuples across the
bus.} encoded by the CAC to $N$ bits (refer to Fig.~\ref{fig:encoder} in
Section~\ref{sec:joint}). Thus, we have
\begin{align*}
    \Rcac & = \frac{K+P}{N}.
\end{align*}
From the equation for $\Recc$ we can solve for $P/N$ as 
\begin{equation*}
    \frac{P}{N} = 1-\Recc.
\end{equation*}
Hence,
\begin{align*}
    \RE
    & = \frac{K}{N} \\
    & = \frac{K+P}{N}-\frac{P}{N} \\
    & = \Rcac+\Recc-1,
\end{align*}
as required.\hfill\IEEEQED

\section{Proof that $\RE \geq \Rs$}
\label{sec:proofs_comparison}

We now show that $\RE(\bm{a}) \geq \Rs(\bm{a})$ for all possible past bus states
$\bm{a}$ with sufficient number of free wires. 

We start by lower bounding $\Rcac(\bm{a})$.  Recall that there are $P$ free
wires on the bus, each of which can carry one information bit. Furthermore,
using the shielding argument seen earlier, we can transmit at least $d/2$ bits
for each alternating run in $\bm{a}$ of any length $d$.  This implies that the
number of information bits that the single CAC can send is at least
\begin{equation*}
    P+(N-P)/2 = (N+P)/2.
\end{equation*}
Hence,
\begin{equation*}
    \Rcac(\bm{a}) \geq (1+P/N)/2.
\end{equation*}

Since $P/N = 1-\Recc$, this implies
\begin{equation}\label{eq:boundRcac}
    \Rcac(\bm{a}) \geq (1+1-\Recc)/2 = 1-\Recc/2.
\end{equation}
From this, we obtain that
\begin{align*}
    \RE(\bm{a})-\Rs(\bm{a})
    & = \Rcac(\bm{a})+\Recc-1-\frac{\Rcac(\bm{a})\Recc}{2-\Recc} \\
    & = \Rcac(\bm{a})\frac{1-\Recc}{1-\Recc/2}+\Recc-1 \\
    & \stackrel{\eqref{eq:boundRcac}}{\geq} 0.
\end{align*}
\hfill\IEEEQED

\section{Proof of Theorem~\ref{thm:dmin}}
\label{sec:proofs_dmin}

Denote by $\Cecc$ and $\Ccac$ the collection of all codewords of the ECC and
CAC, respectively.  We have dropped the dependence of the CAC on the past bus
state $\bm{a}$ from the notation, but we stress that $\Ccac$ depends on
$\bm{a}$. Recall the notation $\de = \de(\bm{a})$ for the minimum distance of the
embedded joint CAC and ECC scheme. Since every codeword of the joint CAC-ECC is
in particular an element of $\Cecc$, we clearly have $\de\geq\decc$. We
next prove the reverse inequality. In the following we use the notation $[N]$ to
denote the set $\{1,2,3,\dots,N\}$.

Let $\mc{P}\subset[N]$ be the location of the parity bits and set $\mc{I}
\defeq [N]\setminus\mc{P}$. For a vector $\bm{c}$ of length $N$, define
$\bm{c}(\mc{I}) \defeq (c_n)_{n\in\mc{I}}$, and similarly define
$\Ccac(\mc{I})$.  Note that, since $\mc{P}$ are
the free wires with respect to $\bm{a}$, we can decide if a vector $\bm{c}$
is in $\Ccac$ by only considering $\bm{c}(\mc{I})$. With slight abuse of
notation, we can therefore write $\bm{c}(\mc{I})\in\Ccac(\mc{I})$ to mean
that $\bm{c}\in\Ccac$.

Let $\bm{c}^{(0)}\in\Cecc$ be a nonzero codeword of minimum weight, i.e.,
$\decc$ by linearity of the ECC. Note that $\bm{c}^{(0)}$ might not be an
element of $\Ccac$.  However, we next prove that we can find
$\bm{c}^{(1)}(\mc{I})\in\Ccac(\mc{I})$ such that
\begin{equation}
    \label{eq:c2}
    \bm{c}^{(2)}(\mc{I}) 
    \defeq \bm{c}^{(0)}(\mc{I})\oplus\bm{c}^{(1)}(\mc{I})\in\Ccac(\mc{I}).
\end{equation}
If this is the case, then we can find the whole codewords $\bm{c}^{(1)}$ and
$\bm{c}^{(2)}$ in $\Cecc$ by computing the necessary parities of
$\bm{c}^{(1)}(\mc{I})$ and $\bm{c}^{(2)}(\mc{I})$ (recall that the ECC is
systematic with the parity bits at locations $\mc{P} = [N]\setminus\mc{I}$). By
construction, we then have $\bm{c}^{(1)},\bm{c}^{(2)}\in\Cecc\cap\Ccac$.
Moreover, by linearity of the ECC, $\bm{c}^{(2)} =
\bm{c}^{(0)}\oplus\bm{c}^{(1)}$.  Hence, the distance between $\bm{c}^{(1)}$ and
$\bm{c}^{(2)}$ is $\bm{c}^{(1)}\oplus\bm{c}^{(2)} = \bm{c}^{(0)}$, which shows
that $\de\leq\decc$, as required.

It remains to construct $\bm{c}^{(1)}(\mc{I})\in\Ccac(\mc{I})$ such that
\eqref{eq:c2} holds. Parse the past bus state $\bm{a}$ into alternating runs,
where each such run is a contiguous subsequence of the form $0101\dots$ or
$1010\dots$.  Recall from Section~\ref{sec:background} that the CAC constraints
are active only within each such run, but not across its boundaries.
Consequently, we can construct $\bm{c}^{(1)}(\mc{I})$ for each such run
independently and then simply concatenate the resulting pieces. In the
following, we will therefore consider, without loss of generality, a past bus
state $\bm{a}$ whose restriction $\bm{a}(\mc{I})$ consists of a single
alternating run, say $0101\dots$.

We explain the procedure of constructing $\bm{c}^{(1)}(\mc{I})\in\Ccac(\mc{I})$
with the help of an example summarized in the following table where each row
corresponds to a vector of $\vert\mc{I}\vert = 11$ bits.
\begin{equation*}
    \begin{array}{c|ccccccccccc}
        \bm{a}(\mc{I}) & 0 & 1 & 0 & 1 & 0 & 1 & 0 & 1 & 0 & 1 & 0 \\
        \bm{c}^{(0)}(\mc{I}) & 1 & 1 & 0 & {\red \bm{0}} & {\red \bm{1}} & {\red \bm{0}} & {\red \bm{1}} & 1 & {\red \bm{1}} & {\red \bm{0}} & 0 \\
        \bm{c}^{(1)}(\mc{I}) & 0 & 0 & 0 & {\red \bm{1}} & {\red \bm{1}} & {\red \bm{1}} & {\red \bm{0}} & 0 & {\red \bm{0}} & {\red \bm{1}} & 0 \\
        \bm{c}^{(2)}(\mc{I}) & 1 & 1 & 0 & 1 & 0 & 1 & 1 & 1 & 1 & 1 & 0 
    \end{array}
\end{equation*}
Observe that we can parse $\bm{c}^{(0)}(\mc{I})$ into positions where the CAC
constraints are satisfied with respect to $\bm{a}(\mc{I})$ and alternating runs
where the constraints are violated (indicated in bold red in the table). At positions
where $\bm{c}^{(0)}(\mc{I})$ satisfies the CAC constraints, we set
$\bm{c}^{(1)}(\mc{I})$ to $0$. Consider then a run of violated constraints in
$\bm{c}^{(0)}(\mc{I})$. Notice that each such run has length at least $2$. If
required, we choose the first and last bit in this run in $\bm{c}^{(1)}(\mc{I})$
such that the CAC constraints are satisfied. The remaining bits in this run in
$\bm{c}^{(1)}(\mc{I})$ are set to $1$. With this construction, there is at least
one value of $1$ in every alternating position of the run of violated CAC
constraints. This is enough to ensure that $\bm{c}^{(2)}(\mc{I}) =
\bm{c}^{(0)}(\mc{I})\oplus\bm{c}^{(1)}(\mc{I})$ satisfies the CAC constraints,
as needed to be shown.  \hfill\IEEEQED

\section{Proof of Theorem~\ref{thm:density}}
\label{sec:proofs_density}

Throughout this section, we make the assumption that the local decoding
neighborhood of each node is a tree. This assumption holds with high
probability asymptotically as $N\to\infty$.

The density evolution for the ECC part of the factor graph is standard. A
message from an information variable node to a ECC-check node is an erasure if and only
if the variable node is erased by the channel, the message from the CAC-check node is an erasure, and the messages from all ECC-check nodes over the
edges other then the current one are erasures. Thus, the probability of erasure is
\begin{equation*}
    \xecc = \varepsilon\ycac\lambda(\yecc).
\end{equation*}
A message from an ECC check node to an information variable node is an erasure
unless all other incoming edges carry non-erasures. The probability of
erasure is
\begin{equation*}
    \yecc = 1-(1-\xp)^2\rho(1-\xecc).
\end{equation*}
Here, the factor $(1-\xp)^2$ arises because each ECC-check
is also connected to two additional parity bits from $\bm{a}_2$ (refer to Fig.~\ref{fig:messages}).

A message from a parity variable node to an ECC-check node is an erasure if the
channel and the other edge coming from the ECC-check are erasures so that
\begin{equation*}
    \xp = \varepsilon \yp.
\end{equation*}
A message from an ECC-check node to a parity variable node is an erasure unless
all its incoming messages are non-erasures so that
\begin{equation*}
    \yp = 1-(1-\xp)R(1-\xecc).
\end{equation*}

Consider next a message from an information variable node to a CAC-check node.
This message is an erasure if and only if the variable node is erased by the
channel and the messages from all ECC-check nodes are erasures.  Hence, the
probability of erasure is
\begin{equation*}
    \xcac = \varepsilon L(\yecc).
\end{equation*}

Consider finally a message from a CAC-check node to an information variable node.
Recall that the probability that this edge is connected
to a CAC-check node of degree $d$ is $\tilde{\rho}_d$. Let $Y$ be the random
variable describing the message over the edge under consideration, and let
$D$ be the random variable describing the degree of the CAC-check node under
consideration. Then the probability of erasure is
\begin{equation}
    \label{eq:ytilde}
    \ycac 
    = \sum_{d=1}^\infty \tilde{\rho}_d\Pp(Y=\,?\mid D = d)
    = 1-\sum_{d=1}^\infty \tilde{\rho}_d\Pp(Y\neq \,?\mid D = d).
\end{equation}

If $d=1$, then the message from this CAC-check node is an erasure with
probability one, so that 
\begin{equation}
    \label{eq:d1}
    \Pp(Y\neq{?}\mid D = 1) = 0. 
\end{equation}

Consider next $d=2$ and assume that the edge under consideration is the first
one. There are two equally likely possible alternating runs in $\bm{a}$ that can
correspond to this CAC check node, $01$ and $10$. Assume for the moment that the
run is $01$. The set of all possible two-bit sequences in $\bm{b}$ satisfying
the CAC constraint is then $\{00, 01, 11\}$. By construction, these three
possible sequences occur with uniform probability in the code ensemble. Note
that $Y$ is a non-erasure if and only if the corresponding variable node has a
value that is forced by its neighboring variable node.  Here, this happens if
the neighboring (second) variable node takes value $0$ and is not erased, which
forces the first variable node to have value $0$ as well. This situation holds
in one of the three possible sequences above, and the probability of non-erasure
is $1-\xcac$. By symmetry, the same reasoning is valid if the edge under
consideration is the second one.  Finally, again by symmetry, the same
conclusion holds if the alternating run in $\bm{a}$ is $10$. Thus, 
\begin{align*}
    \Pp(Y\neq{?}\mid D = 2) 
    & = \frac{1}{3}(1-\xcac) \\
    & = \frac{1}{2}\cdot2\frac{F(1)}{F(4)}(1-\xcac) \\
    & = \frac{1}{2}\sum_{i=1}^2p_{2,i}(\xcac).
\end{align*}

Consider then $d=3$.  There are again two equally likely  possible past bus
states, $010$ and $101$. Assume for now that we are in the first case. The set
of all possible three-bit sequences in $\bm{b}$ satisfying the CAC constraints
are $\{000, 010, 011, 110, 111\}$, and each of them has probability $1/5 =
1/F(5)$ by construction. 

Assume that the edge under consideration is the first one (i.e., $i=1$). For
the first variable node to be forced, the second variable node has to take
value $0$. This happens for one out of the five possible sequences, namely
$000$. Thus, for $i=1$, the probability of $Y$ being a non-erasure is 
\begin{equation*}
    \frac{1}{5}\cdot(1-\xcac) 
    = \frac{F(2)}{F(5)}\cdot(1-\xcac) 
    = p_{3,1}(\xcac).
\end{equation*}
By an analogous argument, for $i=3$, the probability of $Y$ being a
non-erasure is 
\begin{equation*}
    \frac{1}{5}\cdot(1-\xcac) 
    = \frac{F(2)}{F(5)}\cdot(1-\xcac) 
    = p_{3,3}(\xcac).
\end{equation*}

For the middle edge ($i=2$), the variable node is forced if either
neighboring variable node takes value $1$. There is one sequence $111$ for
which both these neighboring variable nodes have to be erased in order for
the current message to be an erasure. There are two messages $\{011, 110\}$
for which only one neighboring variable node has to be erased in order for
the current message to be an erasure. Thus, for $i=2$, the probability of
$Y$ being a non-erasure is 
\begin{equation*}
    \frac{2}{5}\cdot(1-\xcac) 
    +\frac{1}{5}\cdot(1-\xcac^2)
    = \frac{F(1)F(2)+F(2)F(1)}{F(5)}\cdot(1-\xcac) 
    +\frac{F(1)F(1)}{F(5)}\cdot(1-\xcac^2)
    = p_{3,2}(\xcac).
\end{equation*}

This argument was for alternating run in $\bm{a}$ of $010$. By symmetry, the
same conclusion holds if the alternating run in $\bm{a}$ is $101$. Since
the edge is in relative position $i$ with probability $1/3$, we thus have
\begin{equation*}
    \Pp(Y\neq{?}\mid D = 3) 
    = \frac{1}{3}\sum_{i=1}^3p_{3,i}(\xcac).
\end{equation*}

Consider finally the case of general $d>3$. Consider first edge position $i=1$,
and without loss of generality consider an alternating run in $\bm{a}$ of
$0101\dots$. There are $F(d+2)$ possible sequences that satisfy the CAC
constraints with respect to this alternating run. In order for the first
variable to be forced to take value $0$, the sequence must be of the form
$000\star\dots$, where $\star$ indicates that the corresponding value can be
either $0$ or $1$. Note that the third value is forced to $0$ to satisfy the CAC
constraints. There are $F(d-1)$ such sequences. Hence, for $i=1$, the
probability of $Y$ being a non-erasure is 
\begin{equation*}
    \frac{F(d-1)}{F(d+2)}\cdot(1-\xcac) 
    = p_{d,1}(\xcac).
\end{equation*}
By an analogous argument, for $i=d$, the probability of $Y$ being a
non-erasure is 
\begin{equation*}
    \frac{F(d-1)}{F(d+2)}\cdot(1-\xcac) 
    = p_{d,d}(\xcac).
\end{equation*}

Consider then $i\in\{2,3,\dots,d-1\}$, and without loss of generality consider
an alternating run in $\bm{a}$ of $\dots101{\red\bm{0}}101\dots$, where $i$ is the
position indicated in bold red font. For the variable node $i$ to be constrained
from both sides, the sequence in $\bm{b}$ has to be of the form
$\dots\star00{\red\bm{0}}00\star\dots$.  There are $F(i-1)F(d-i)$ such sequences. For the
variable node $i$ to constrained from only the left, in $\bm{b}$ has to be of
the form $\dots\star00{\red\bm{0}}1\star\star\dots$. There are $F(i-1)F(d-i+1)$ such
sequences.  For the variable node $i$ to be constrained from only the right, the
sequence in $\bm{b}$ has to be of the form $\dots\star\star1{\red\bm{0}}00\star\dots$.
There are $F(i)F(d-i)$ such sequences. Hence, for $i\in\{2,3,\dots,d-1\}$, the
probability of $Y$ being a non-erasure is 
\begin{equation*}
    \frac{F(i-1)F(d-i+1)+F(i)F(d-i)}{F(d+2)}\cdot(1-\xcac) 
    +\frac{F(i-1)F(d-i)}{F(d+2)}\cdot(1-\xcac^2)
    = p_{d,i}(\xcac).
\end{equation*}

These probabilities do not depend on the actual realization of the alternating
run in $\bm{a}$ other than its length. Since the edge is in relative position
$i$ with probability $1/d$, we thus have
\begin{equation*}
    \Pp(Y\neq{?}\mid D = d) 
    = \frac{1}{d}\sum_{i=1}^dp_{d,i}(\xcac).
\end{equation*}

Substituting this expression and \eqref{eq:d1} into~\eqref{eq:ytilde} yields
that
\begin{equation*}
    \ycac 
    = 1-\sum_{d=2}^\infty \frac{\tilde{\rho}_d}{d}
    \sum_{i=1}^dp_{d,i}(\xcac).
\end{equation*}
This concludes the proof. \hfill\IEEEQED

\end{appendices}

\newcommand{\SortNoop}[1]{}

\end{document}